# Investigation of structure and hydrogen bonding of super-hydrous phase B (HT) under pressure using first principles density functional calculations

**H. K. Poswal, Surinder M Sharma and S. K. Sikka**[†]

High pressure Physics Division, Bhabha Atomic Research Centre, Mumbai 400085, India

[†] Office of the Principal Scientific Adviser to the Government of India, Vigyan Bhawan Annexe, New Delhi 110011, India

**Abstract**

High pressure behaviour of superhydrous phase B(HT) of $Mg_{10}Si_3O_{14}(OH)_4$ (Shy B) is investigated with the help of density functional theory based first principles calculations. In addition to the lattice parameters and equation of state, we use these calculations to determine the positional parameters of atoms as a function of pressure. Our results show that the compression induced structural changes involve cooperative distortions in the full geometry of the hydrogen bonds. The bond bending mechanism proposed by Hofmeister et al [1999] for hydrogen bonds to relieve the heightened repulsion due to short H---H contacts is not found to be effective in Shy B. The calculated O-H bond contraction is consistent with the observed blue shift in the stretching frequency of the hydrogen bond. These results establish that one can use first principles calculations to obtain reliable insights into the pressure induced bonding changes of complex minerals.

# 1. Introduction

Several recent investigations of geophysical interest indicate that the minerals in upper mantle and transition zone may be substantially hydrated. High pressure experiments suggest that dense hydrous magnesium silicates could be stable under geothermal conditions along the subduction slabs. Amongst these, the superhydrous phase B (Shy-B) has been viewed as one of the most important minerals responsible for water transport to almost the top of the lower mantle. The stability fields of these minerals has been investigated using various high pressure experiments, particularly with the help of x-ray diffraction measurements. However, x-ray diffraction method does not provide hydrogen atom positions accurately and always unambiguously. This information is essential to investigate the changes in the hydrogen bonding in these minerals at high pressures. Advances in the first principles density functional calculations and availability of higher computer power permit now to reliably determine the position of atoms, including that of hydrogen atom, in the crystal structures of complex minerals. In this paper, we use the first principles density functional calculations to study the pressure induced evolution of the structure and hydrogen bonding in superhydrous phase B of magnesium silicate.

At 0.1 MPa, the hydrous mineral phases, Phase A ($Mg_7Si_2O_6(OH)_6$), Phase B ($Mg_{12}Si_4O_{19}(OH)_2$) and superhydrous Phase B ($Mg_{10}Si_3O_{14}(OH)_4$) contain two hydroxyl groups, hydrogen bonded to the same acceptor oxygen atom [Kagi et al., 2000; Finger at el., 1991; Pacalo and Parise, 1992]. This arrangement, shown in Fig. 1, is very unusual.

The angle H---O---H is ~ 60°. The non bonded H---H contacts at 0.1 MPa, as determined by Philips et al [1996] by NMR technique are 2.10 Å, 1.86 Å and 1.83 Å in Phase A, Phase B and superhydrous phase B (Shy B) respectively. These are smaller than the value 2.4 Å, which is twice the van der Waals radius of the hydrogen atom and thus produce a repulsive strain in the structure. Application of pressure should further decrease these distances and intensify this repulsive strain. A hydrogen bond, which is a relatively weak interaction, may counter the effect of short H---H distances by compression of the O-H distance, bending of the hydrogen bond and disordering of the hydrogen atom sites [Liu et al., 2003]. For Phase B, Hofmeister et al [1999] have proposed the hydrogen bond bending as the strain relieving mechanism, where the H-O---O angles for the two hydrogen bonds were postulated to increase by $9 \pm 1^0$ and $11 \pm 1^0$ from 0 to 37 GPa. For Phase A, Poswal et al [2009] have shown through density functional calculations that the full geometry of the hydrogen bonds is co-operatively involved i.e., the hydrogen bond bending by $3.3^0$ and $2.6^0$ for the two hydrogen bonds, opening of the H---O---H angle by $8^0$ and a slight decrease ( ~ 0.04Å) of one of the O-H bond lengths from 0 to 45 GPa. To throw more light on the H---H strain relieving mechanisms in these compounds, we have now carried out first principles DFT calculations on superhydrous Phase B.

The stability of Shy B has been studied by many authors [see Litasov et al., 2007, and references therein] and it is now believed that if it exists inside the earth, it will be stable up to bottom of the transition zone and top of the lower mantle, within the colder subducted slabs. However, there has been some controversy about the crystal structure of Shy B. Synthesized at 20 GPa and 1673K, Pacalo and Parise [1992] found that Shy B crystallizes in the orthorhombic space group Pnnm with one independent hydrogen atom. However, Kudoh et al [1994] found a non-centrosymmetric structure in space group P2$_1$mn with two non-equivalent hydrogen atoms. Later spectroscopic [ Cynn et al. ,1996; Frost and Fei, 1998; Hofmeister et al., 1999 and Liu et al., 2002] and NMR studies by Philips et al [1996] were consistent with the structure containing two independent hydrogen positions. However, recently, Koch-Muller et al [2005] have shown that the crystal structure of Shy B is dependent on the synthesizing conditions. Shy B synthesized at 22 GPa and 1673 K is centrosymmetric with space group Pnnm, the same as that of Pacalo and Parise [1992], whereas the one synthesized at 22 GPa and 1473 K or below occurs in the non-centrosymmetric space group Pnn2. which is different from that determined by Kudoh et al [1994]. They denote these two forms as HT and LT polymorphs of Shy B. Because of the available computational power constraints, we have done DFT studies for the 70 atoms/unit cell HT form only as it has 30 variables (3 unit cell parameters + 27 independent coordinates) in its asymmetric unit compared to 57 for the LT form.

The P-V-T relations have been measured for Shy B up to about 30 GPa and 1800 K [see Litasov et al., 2007, and references therein]. However, we notice that none of these are on samples of HT form. Only Pacalo and Weidner [1996] measured the sound velocities at ambient conditions on the HT sample and derived the isothermal bulk modulus. The infrared studies up to 16.4 GPa at room temperature on HT Shy B, by Koch-Muller et al [2005], indicate that the O-H stretching frequency is slightly blue shifted ($\delta\nu/\delta P$= + 0.282 cm$^{-1}$/GPa) even though $\nu_0$ is 3373 cm$^{-1}$, where it should have normally red shifted [see Sikka, 2007]. We also offer an explanation for this behaviour from our present studies.

## 2. Computational Details

First principles structure relaxation calculations were performed using the density functional theory (DFT) within the frame work of projector augmented wave method as implemented in the VASP (Vienna *ab initio* simulation package) code [Kresse et al., 1994, 1996]. These calculations were carried out on *Anupam Ajeya* supercomputer at Bhabha Atomic Research Centre up to 30 GPa. Constant volume variable cell shape simulations were done by explicitly treating two valence electron of Mg ($2s^22p^0$), four of Si ($2s^22p^2$), six of O ($2s^22p^4$), one of H ($1s^1$). We utilised the Perdew Burke Ernzerhof generalised gradient approximation (GGA) for exchange and correlation functional [Perdew et al., 1996]. This functional has earlier been very successful in calculations of structural properties of hydrogen-bonded systems [Hammann, 1997]. To confirm this, we carried out some calculations using both GGA and local density approximation (LDA). At the experimental volume of 618.16Å$^3$, the calculated pressure comes out to be 4.5 GPa for GGA and −4.68 GPa for LDA. This level of discrepancy between GGA and LDA is common with some other calculations [see for example, Panero and Stixrude, 2004] indicating that GGA generally over-estimates and LDA under-estimates the lattice volume. Also, for atomic positions and bond parameters, GGA calculations gave results in closer agreement with those of the experiment. Several more calculations were also performed using hard potential of Si, O, H. In these cases Mg $1p^62s^2$ electrons were treated explicitly. However, these calculations did not improve the agreement with experiments. Brillouin zone was sampled by the Monkhorst–Pack scheme [Monkhorst and Pack,1976]. Before undertaking the detailed calculations, we tested for the convergence with respect to the number of plane waves in the basis set and for **k** points sampling in the Brillouin zone. The calculations presented here were carried out with 6 × 4 × 4 **k** point mesh in the whole Brillouin zone( energy convergence < 0.003meV), and the basis set was expanded by taking plane wave cutoff as 600 eV which gives energy convergence of < 0.047meV/atom. The calculated Hellman–Feynman forces were converged, until the largest force component was less than $1 \times 10^{-3}$ eV/Å. Calculations were started from the ambient structure determined by Pacalo and Parise [1992].

3. **Results and Discussion**

Table 1 shows the calculated unit cell parameters at 0.1 MPa along with the experimental values. At 0.1 MPa, these parameters agree with each other to about 1%. This is similar to that for some other minerals [see for example Winkler et al., 2008 for zoisite]. However, at the experimental volume at ambient pressure, also given in Table 1, the cell parameters are in excellent agreement with each other. Therefore, in order to make meaningful comparisons, the theoretical and experimental atomic coordinates [Pacalo and Parise,1992] are given in Table 2 at the same volume. We found that the LDA calculations predicted nearly identical coordinates at the same volume. The level of agreement for non hydrogen atoms can be classified as excellent. The larger differences for hydrogen atom are not unexpected. This is because the experimental values are from X-ray diffraction measurements, while the theoretical values are for ionic positions. Some bond lengths and angles are compared in Table 3. The differences in bond lengths are mostly less than 0.01Å. This level of agreement gives confidence that the results on HT Shy B at high pressure will be reliable.

The calculated bulk modulus from a third order Birch-Murnaghan [Birch, 1978] fit to the calculated pressure volume data is 137.5 GPa at the theoretical equilibrium volume of 638.9 Å$^3$. The only value with which it can be compared is the one derived by Palco and Weidner [1996] from their measured acoustic velocities at ambient conditions.

This value is 154 ±4 GPa. However, this is at the volume of 618.2 Å$^3$. Correcting the theoretical bulk modulus for this difference in volumes, by using our DFT determined pressure derivative of 4.27, the obtained value of 156.9 GPa becomes consistent with the experimental value.

Based on the high pressure IR measurements, Koch-Muller et al [2005] have mentioned that the HT form is more compressible compared to the LT Shy B. To check this, we have plotted the experimental data for LT Shy B against our theoretical curve for HT Shy B in Fig. 2 (It may again be emphasized that no experimental data for HT Shy B is available). There is good agreement to about 15 GPa, with the theoretical curve becoming a bit softer beyond that pressure. This level of agreement is not surprising as the space group of the LT Shy B is a sub group of the HT Shy B [ Koch-Muller et al., 2005]. In such cases, one expects that the equations of state should not differ much from each other [Sikka et al., 1998 ]. Same consistency is obtained for the axial ratios as illustrated in Fig. 3. All the above is supported by a recent NMR experiments on LT Shy B by Xue et al [2008]. They find that the structure of LT Shy B in space group Pnn2 deviates very slightly from the higher-symmetry space group (Pnnm) of HT Shy B.

The hydrogen bond parameters as a function of compression are presented in Table 4. Contrary to the expectation from the well established correlation between O-H versus H---O (or O---O) lengths, the O-H length in HT Shy B deceases slightly with compression. This is very similar to our finding for one of the O-H distances in Phase A. This is very clearly illustrated in Fig. 4. This departure in opposite direction from the correlation is, perhaps, the first one calculated so far although Friedrich et al [2007] have calculated by DFT a slower increase of O-H against H---O distance in diaspore under pressure.

The angle H---O3---H also increases by about 2º . The value of 1.79 Å for the non-bonded H---H contact in neighboring hydrogen bonds is close to the value 1.83 Å , determined by Philips et al [1996]. This further decreases with compression. According to Hofmeister et al [1999], in analogy with Phase B, the enhanced repulsion due to contraction of the non-bonded H---H distance with compression should substantially increase the hydrogen bond bending angle i.e. an increase in the H-O5---O3 angle here. Instead, we find that this angle first decreases and only after 15 GPa, it begins to increase again (Table 4). However, this decrease is much smaller. Combining this with the results for Phase A, we can now definitely say that the hydrogen bond bending mechanism for relieving the heightened repulsion due to short H---H contact, as proposed by Hofneister et al, is not that effective in HT Shy B.

The H---H distance of 1.79Å at 0.1 MPa is near its limiting value of about 1.8 Å [Sikka and Sharma, 2008] and it should have already resulted in transition to another phase. However, no such transition has been detected so far in HT Shy B, as shown by infrared measurements up to 16.4 GPa by Koch-Muller et al [2005]. In this context, the existence of two polymorphs, HT and LT, in Shy B may be noted. However, the relationship between the two polymorphs under pressure has not yet been investigated. It may also be pointed out that HT Shy B is metastable at 0 K at which these calculations are performed and in some cases the range of metastability may be large. For example, recently, Friedrich et al [2007] have reported that α-AlOOH (diaspore) persists up to ~ 50 GPa at room temperature. But high temperature experiments [Suzuki et al., 2000; Ohtani et al., 2001] show that about 17 GPa it transforms to δ-AlOOH, in excellent agreement with the GGA transition pressure of ~ 18 GPa [Li et al., 2006].

The calculated contraction in O-H is also in accord with the small blue shift for stretching frequency for HT Shy B observed by Koch-Muller et al [2005]. According to the correlation between the frequency of the O-H stretch mode versus assembled by Libowitzky [1999] from data on different minerals at 0.1 MPa, $v_0$ = 3373 cm$^{-1}$ should have shifted to ~ 2800 cm$^{-1}$ for H---O distance variation up to 16 GPs in experiment ( we assume H---O distances from Table 4 to be valid here). Instead, it was found to shift in opposite direction with $\delta v/\delta P$ of 0.282 cm$^{-1}$/GPa. The magnitude of decrease in O-H length (0.003Å up to 30 GPa) is of the same order as given by Alabugin et al [2003] for blue shifted O-H---O bonds The observed small positive shift of the O-H stretching frequency is also in accord with that given by Fan et al [2002].

4. Conclusions

Our first principles calculations provide a clear picture for strain relieving mechanism for H---H repulsion in Shy B. Again, like for Phase A, this involves the change of full hydrogen bond geometry under pressure. This is different from the bond bending mechanism proposed by Hofmeister et al for Phase B. When the available computing power will substantially increase in near future, we intend to test this directly for Phase B.

**Figure Captions**

**Figure 1:** Configuration of two hydroxyl groups in superhydrous phase B, hydrogen bonded to the same acceptor oxygen atom.

**Figure 2**: $V/V_0$ as a function of pressure. Solid line represents the results of present calculations. Solid square, open circle, open triangle represent the experimental data from Litasov et al [2007], Crichton et al [1999] and Shieh et al [2000] respectively for LT Shy B.

**Figure 3**: Variation of normalized lattice parameters as a function of pressure. Solid line represents the results of present calculations. Solid squares, open circles, open triangles represent the observed experimental data from Litasov et al [2007], Crichton et al [1999] and Shieh et al [2000] respectively for LT Shy B.

**Figure 4**: (Colour online) Computed O-H distances for Phase A and Shy B versus the correlation of O-H and H---O hydrogen bond lengths in some inorganic compounds and minerals assembled from 0.1 MPa neutron diffraction data from different compounds [taken from Sikka and Sharma, 2008]

Table 1: Comparison of calculated and observed lattice parameters.

| | Calculation[a] | Calculation[b] | Experiment[c] | Experiment[d] |
|---|---|---|---|---|
| $V_o(Å^3)$ | 638.95 | 618.16 | 618.16(10) | 621.48 |
| A (Å) | 5.149 | 5.0913 | 5.0894(6) | 5.0977(1) |
| B (Å) | 14.091 | 13.9403 | 13.968(7) | 13.991(3) |
| C (Å) | 8.8005 | 8.7097 | 8.6956(2) | 8.718(1) |

[a] By GGA for P=0

[b] By GGA for experimental volume at P=0.1 MPa

[c] from Pacalo and Parise (1992)

[d] from Koch-Muller et al (2005) for composition $Mg_{8.32} Ni_{1.77} Si_{2.95} O_{14}(OH,D)_4$

Table 2: The comparison between the observed and computed fractional coordinates of atoms for HT shy B at the ambient experimental volume of ~ 618.16Å$^3$.

| Atom | coordinate | Ambient volume 618.16 Å$^3$ | |
|---|---|---|---|
| | | Experimental[a] | Calculated ($P_{calc}$ = 4.5 GPa) |
| Si1 | | at special position (0.5,0.0) | |
| Si2 | x | 0.4860(8) | 0.4853 |
| | y | 0.3765(3) | 0.3764 |
| Mg1 | x | 0.1694(10) | 0.1668 |
| | y | 0.1735(4) | 0.1742 |
| Mg2 | x | 0.1583(7) | 0.1576 |
| | y | 0.3234(3) | 0.3231 |
| | z | 0.3238(4) | 0.3240 |
| Mg3 | z | 0.3212(6) | 0.3219 |
| Mg4 | z | 0.3418(6) | 0.3413 |
| O1 | x | 0.3322(14) | 0.3315 |
| | y | 0.4134(5) | 0.4140 |
| | z | 0.1561(8) | 0.1574 |
| O2 | x | 0.4930(22) | 0.4919 |
| | y | 0.2592(7) | 0.2579 |
| O3 | x | 0.7914(21) | 0.7964 |
| | y | 0.4145(7) | 0.4136 |
| O4 | x | 0.3564(14) | 0.3541 |
| | y | 0.0730(5) | 0.0738 |
| | z | 0.1446(8) | 0.1461 |
| O5 | x | -0.0174(18) | -0.0192 |
| | y | 0.2518(5) | 0.2533 |
| | z | 0.1592(8) | 0.1584 |
| O6 | x | 0.1956(20) | 0.1942 |
| | y | -0.0792(7) | -0.0794 |
| H | x | -0.0769(37) | -0.0801 |
| | y | 0.3026(13) | 0.3104 |
| | z | 0.1085(17) | 0.1014 |

[a] from Pacalo and Parise (1992)

Table 3. Some selected interatomic distances and angles of Shy B

| Ambient volume 618.16 Å$^3$ | | |
|---|---|---|
| Bond lengths and angles | Experimental[a] (Å) | Calculated (Å) ($P_{calc}$ = 4.5 GPa) |
| Si1-O4 | 1.7770(7) | 1.7973 |
| -O6 | 1.903(1) | 1.9106 |
| Si2-O2 | 1.639(1) | 1.6517 |
| -O3 | 1.6423(12) | 1.6672 |
| -O1 | 1.6486(7) | 1.6641 |
| Mg1-O5 | 2.0045(9) | 2.0039 |
| -O2 | 2.0365(13) | 2.0253 |
| -O4 | 2.1109(9) | 2.1177 |
| -O6 | 2.2776(12) | 2.2633 |
| Mg2-O5 | 1.9620(9) | 1.9593 |
| -O5 | 1.9625(10) | 1.9655 |
| -O2 | 2.0949(9) | 2.0824 |
| O1 | 2.1166(9) | 2.1203 |
| -O4 | 2.1272(10) | 2.1261 |
| -O6 | 2.1790(9) | 2.1832 |
| Mg3-O1 | 2.0645(9) | 2.0552 |
| -O4 | 2.1023(8) | 2.0948 |
| -O6 | 2.1521(8) | 2.1469 |
| Mg4-O1 | 2.0808(8) | 2.0701 |
| O3 | 2.1080(9) | 2.1056 |
| O4 | 2.1255(9) | 2.1217 |
| O5-H1 | 0.8879(175) | 0.9881 |
| O3---H1 | 1.9448(175) | 1.8018 |
| ∠H1---O3---O5) | 0.236° | 0.761° |
| ∠ (H1---O3---H1) | 58.040° | 58.677° |

[a] from Pacalo and Parise (1992)

Table 4: Computed hydrogen bond parameters for HT Shy B as function of pressure

| GGA Pressure (GPa) | V/Vo | O5---O3 | O3---H1 | O5-H1 | H---H | ∠(H1-O5---O3) (degrees) | ∠H1---O3---H1 (degrees) |
|---|---|---|---|---|---|---|---|
| -0.57 | 1.002 | 2.8312 | 1.8431 | 0.9886 | 1.7914 | 1.500 | 58.153 |
| 4.55 | 0.967 | 2.7894 | 1.8018 | 0.9881 | 1.7656 | 1.394 | 58.723 |
| 9.59 | 0.939 | 2.7546 | 1.7674 | 0.9876 | 1.7442 | 1.312 | 59.133 |
| 14.39 | 0.916 | 2.7258 | 1.7392 | 0.9870 | 1.7268 | 1.287 | 59.529 |
| 21.82 | 0.884 | 2.6875 | 1.7018 | 0.9862 | 1.7038 | 1.325 | 60.079 |
| 26.04 | 0.869 | 2.6684 | 1.6832 | 0.9856 | 1.6923 | 1.367 | 60.359 |
| 29.22 | 0.858 | 2.6548 | 1.6701 | 0.9853 | 1.6843 | 1.415 | 60.563 |

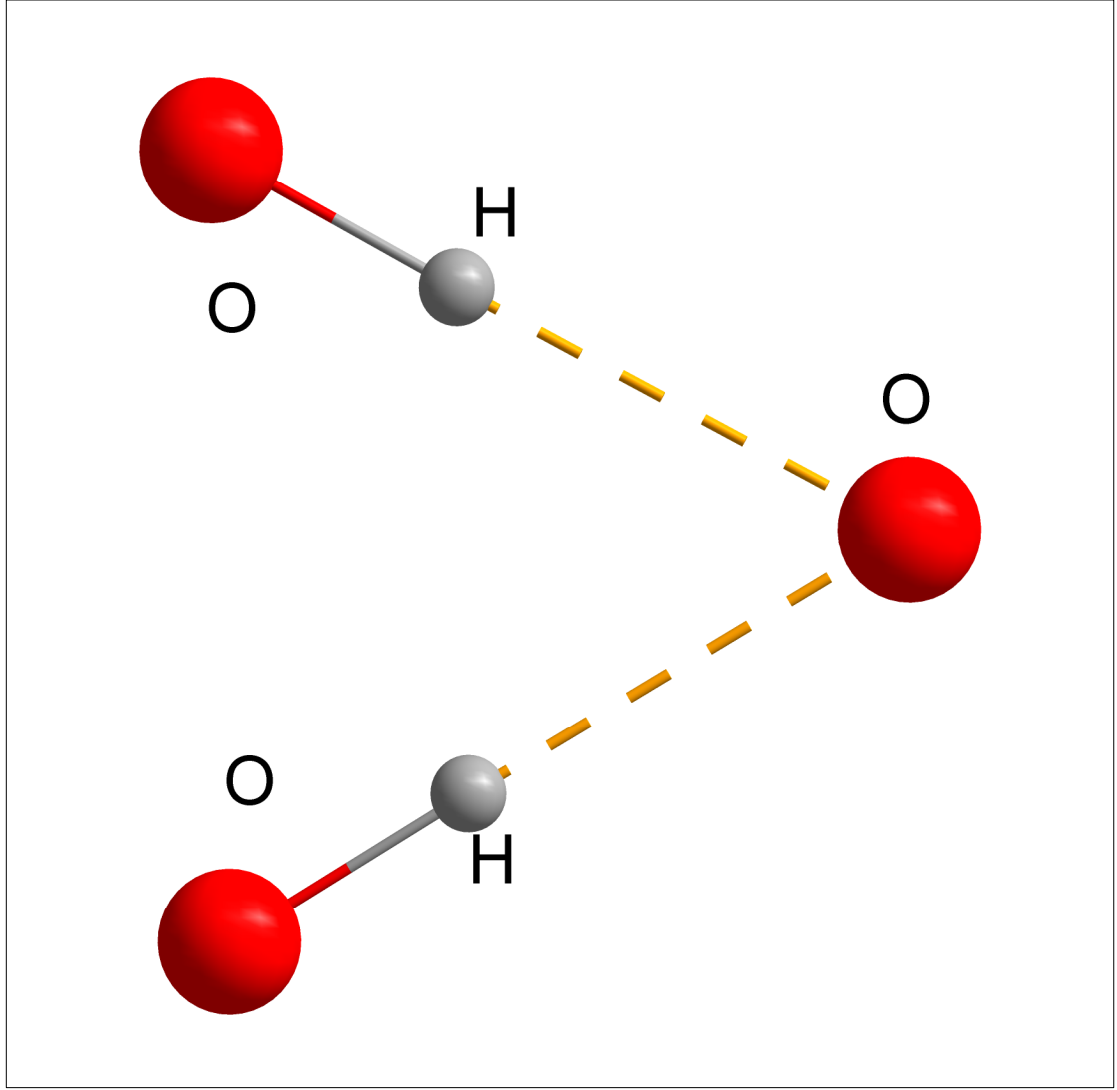

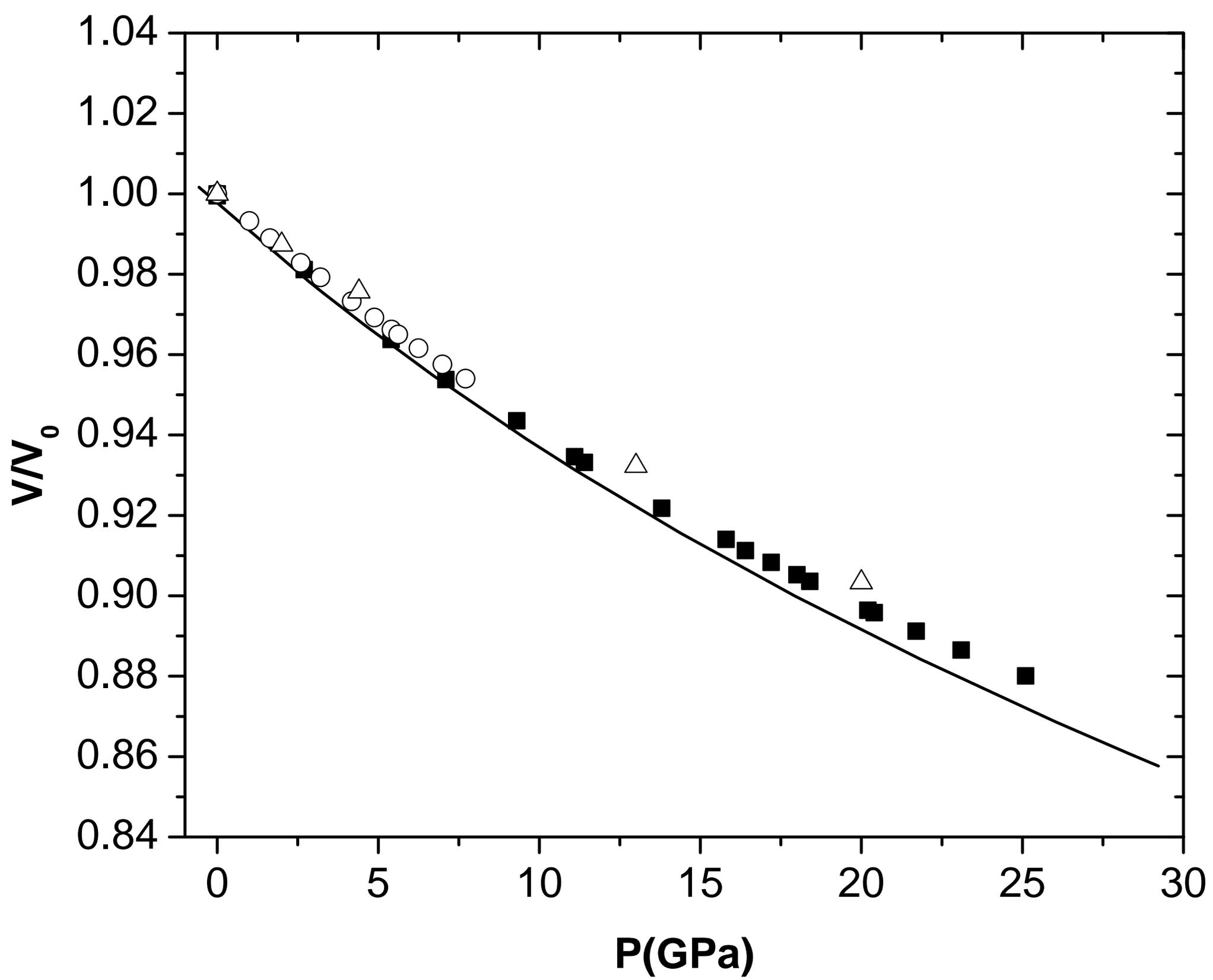

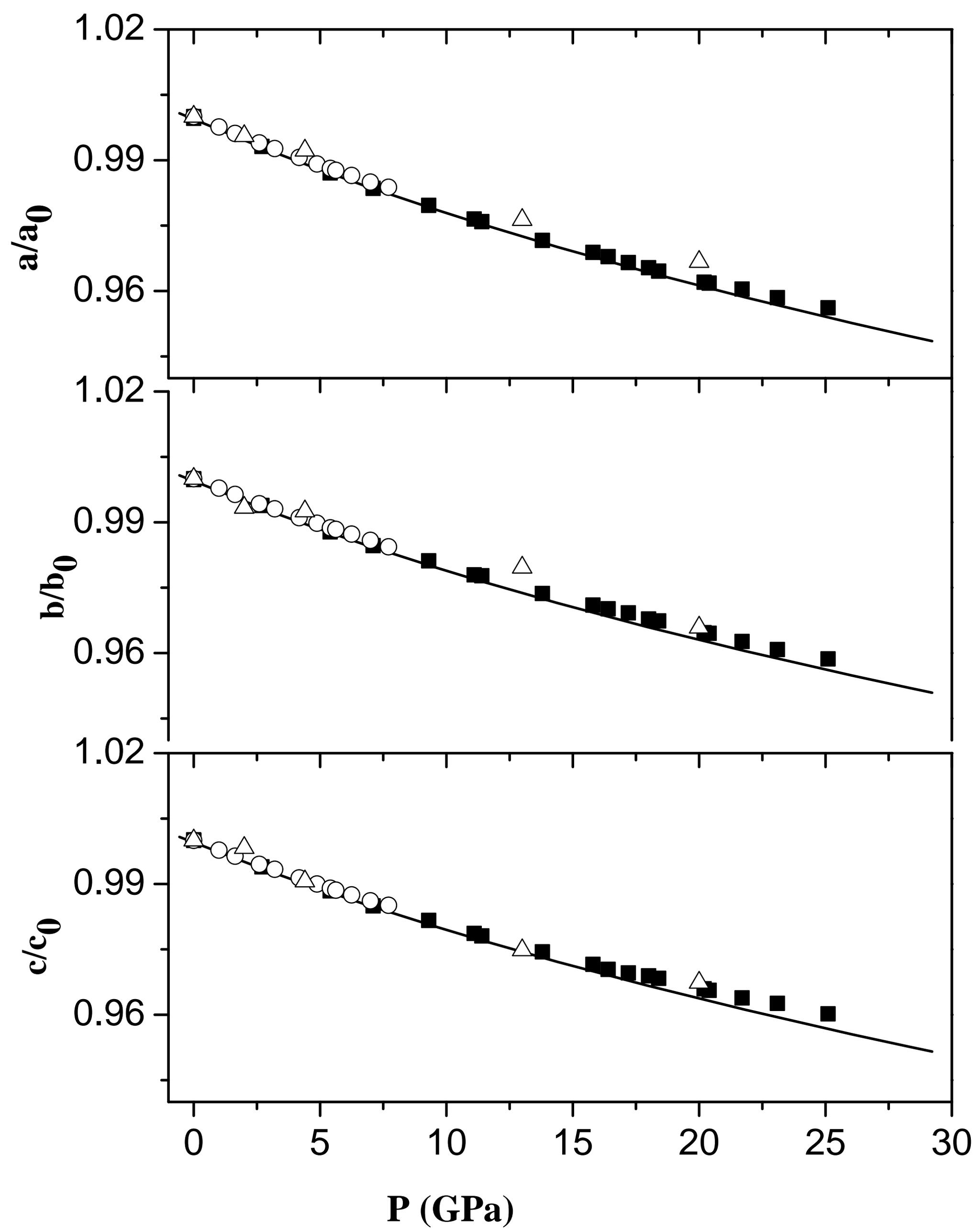

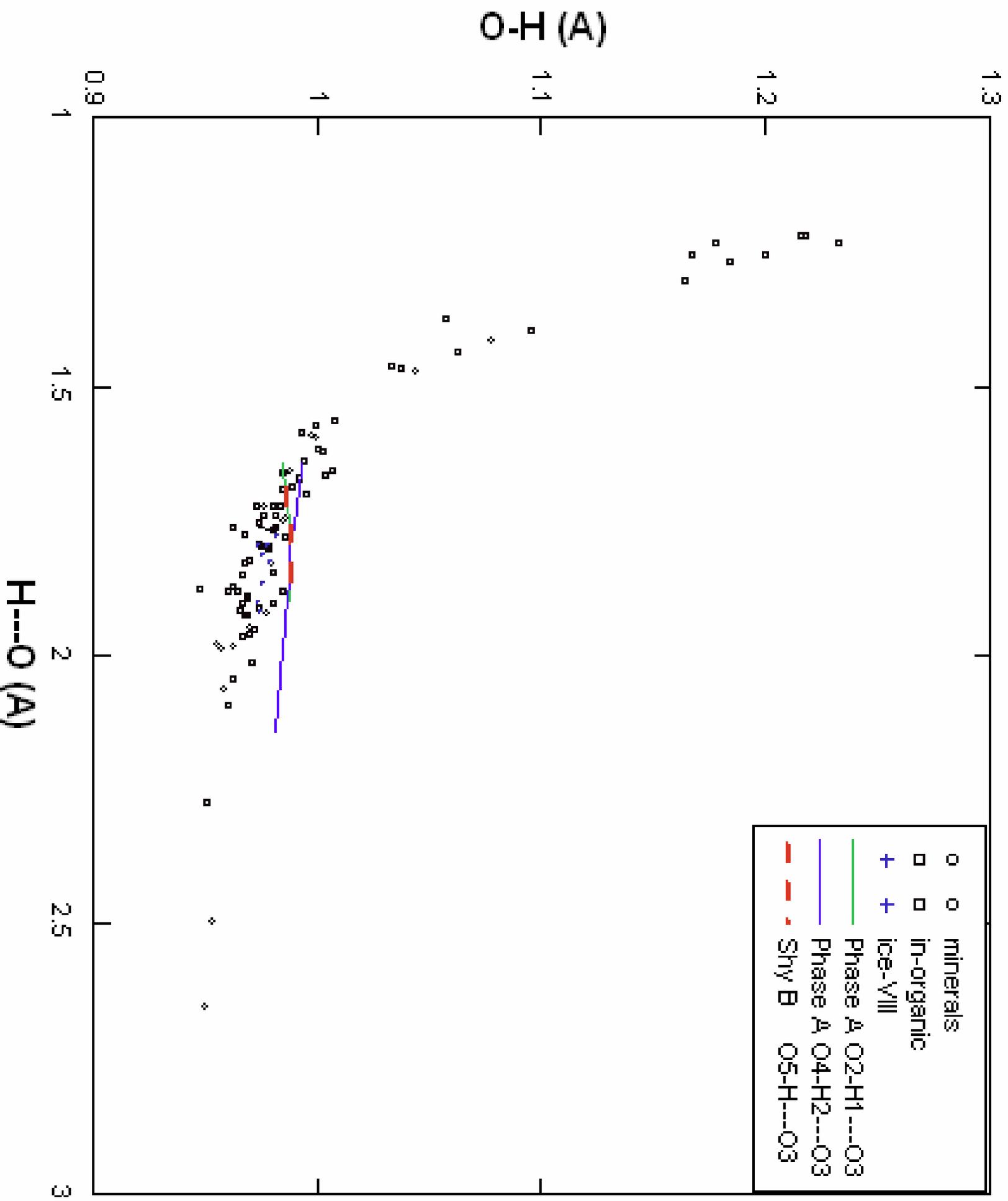